# Enhanced Thermal Conduction through Nanostructured Interfaces

*Woosung Park,[1] Aditya Sood,[1,2] Joonsuk Park,[2] Mehdi Asheghi,[1] Robert Sinclair,[2]*

*and Kenneth E. Goodson[1*]*

[1]Department of Mechanical Engineering, [2]Department of Materials Science and Engineering,

Stanford University, Stanford, California 94305, USA

**ABSTRACT**

Interfaces dominate heat conduction in nanostructured systems, and much work has focused on methods to enhance interfacial conduction. These approaches generally address planar interfaces, where the heat flux vector is everywhere normal to the interface. Here, we explore a nanostructured interface geometry that uses nonplanar features to enhance the effective interfacial conductance beyond what is possible with planar interfaces. This interface consists of interdigitating Al pillars embedded within $SiO_2$ with characteristic feature size ranging from 100 nm to 800 nm. The total sidewall surface area is modulated to highlight the impact of this additional channel by changing the pillar-to-pillar pitch $L_P$ between 1.6 $\mu$m and 200 nm while maintaining the same Al:$SiO_2$ fill fraction. Using optical pump-probe thermoreflectance measurements, we show that the effective conductance of a ~65 nm thick fin layer monotonically increases with decreasing $L_P$, and that the conductance for $L_P$ = 200 nm is more than twice the prediction for a layered stack with the same volume ratio and a planar interface. Through a combination of Boltzmann transport modeling and finite element calculations, we explore the impact of the pitch $L_P$ and the pillar aspect ratio on effective thermal conductance. This analysis suggests that the concept of nanostructured interfaces can be extended to interfaces between diffusive and quasi-ballistic media in highly scaled devices. Our work proposes that the controlled texturing of interfaces can facilitate interfacial conduction beyond the planar interface regime, opening new avenues for thermal management at the nanoscale.

**INTRODUCTION**

Interfaces strongly influence thermal transport in nanostructured systems and have become a primary impediment to thermal management in many applications such as nanoelectronics [1, 2], energy conversion devices [3-5], and nanophotonics [6-8]. Much research has targeted an improved understanding of the impact on the thermal interfacial resistance of the phonon density-of-states mismatch [9, 10], near interfacial defects such as vacancies and dislocations [11-15], interface roughness [16-18], phonon inelastic scattering [19-21], electron-phonon interactions [22-25], among other factors. Current approaches to enhancing interfacial thermal transport rely primarily on improving interface quality, accomplished by reducing surface roughness and mitigating near-interfacial defects [21, 26, 27], or through the introduction of vibrational and lattice-matching interlayers [17, 28]. While the specifics vary, the unifying theme among existing approaches is that they involve planar interfaces, where the heat flux vector is everywhere parallel to the surface normal. In order to push the limits with planar interfaces, new approaches are required that can enhance thermal transport beyond what is currently possible.

In this work, we explore an interface geometry that comprises non-planar features with characteristic length scales down to 100 nm. Through the creation of nano-fabricated fin-like projections, we increase the total area of contact and make the interfacial heat flow three-dimensional in localized regions. This nonplanarity increases the area for heat flow, which enhances the effective interfacial conductance. Our approach is inspired by macro-scale fin arrays that are conventionally used to enhance heat transfer between solid surfaces and fluids by increasing contact area [29]. Previous research has explored the use of micro-structured surfaces for thermal interface materials at the chip packaging level [30, 31], and computational studies have suggested that the extension of this idea to atomic length scales can be beneficial to interfacial conduction [32]. However, the demonstration of nanostructured interfaces between solids in intimate contact, *e.g.* a stack of layered thin films, has proved challenging so far. Here, we demonstrate that a significant enhancement in thermal conduction is possible in a nanoscale solid-state fin array when the fin-pitch becomes comparable to the thermal healing length along the layer of nanostructured interfaces.[33] Using Boltzmann transport modeling and finite element simulations, we show that the thermal conduction across sidewall interfaces plays an increasingly important role in the conduction across structured interfaces as fin dimensions decrease. We propose that controlled nonplanarity can be useful for enhancing the effective thermal conductance of an interface between any pair of solid materials with sufficient contrast in thermal conductivity and comparable magnitude of volumetric and interfacial resistances.

**THERMAL CHARACTERIZATION OF NANOSTRUCTURED FIN ARRAY**

We explore the concept of nanostructured interfaces experimentally by fabricating a nanostructured fin array (NFA) between adjacent layers of Al and $SiO_2$. The fin array comprises a layer of interdigitating teeth-like structures extending between Al and $SiO_2$, with cuboidal pillars of Al embedded within the $SiO_2$ in the form of a periodic square lattice, as shown in Fig. 1. The pillars have a nominal thickness of ~65 nm, and a constant Al:$SiO_2$ fill fraction of 1:3. The square lattice has a pitch $L_P$ that varies between 200 nm and 1.6 $\mu$m. With decreasing $L_P$ (while maintaining constant thickness), the aspect ratio of Al pillars increases, which leads to increased contact area at sidewall interfaces. Direct contact between the Al pillars and the top contiguous Al layer used for thermoreflectance measurements ensures that this configuration effectively acts to extend the area of contact between adjacent layers of Al and $SiO_2$ through the creation of finger-like extensions at the interface (see Supplementary Information for sample fabrication details). At constant fill fraction, as $L_p$ decreases, a larger fraction of the heat flux from the Al to SiO$_2$ flows through the Al pillars, causing an increase in the effective conductance of the structured interface. We compare the effective thermal conductivity of the nanostructured fin layer with that of the reference stack, which is defined as a layered structure with the same Al:SiO$_2$ volume ratio and Al-SiO$_2$ thermal interfacial resistance with no nanostructured fins present.

### Time-domain Thermoreflectance

We measure the effective thermal conductivity of the nanostructured fin array layer in the film-normal direction, $\kappa_{NFA}$, using time-domain thermoreflectance (TDTR), an optical pump-probe technique [11, 34, 35]. TDTR technique utilizes ~ 9 ps optical pulses from a pump source to heat up the surface of a sample coated with a thin Al transducer layer, and time-delayed probe pulses to interrogate the temporal changes in transducer reflectivity as heat diffuses into the sample of interest (see Supplementary Information for the details of TDTR). A schematic of our multilayer sample stack is shown in Fig. 2(a). We treat the stack as comprising three layers on top of a semi-infinite silicon substrate: (from top to bottom) the Al transducer, the nanostructured fin array layer, and SiO$_2$. The fin array layer is assumed to be laterally homogeneous, and isotropic, although we are insensitive to its in-plane conductivity because the thermal penetration depth (~250 nm at 4 MHz) is much smaller than the laser spot size (~10 $\mu$m). We assume literature values for the thermal conductivity of the SiO$_2$ and Si [36], and for the volumetric specific heat of the Al [37], SiO$_2$ [38] and Si [39]. To determine the thermal conductivity of the Al transducer layer, we perform four probe measurements of electrical conductivity of a patterned metal line on a $SiO_2$/Si substrate, which is metallized at the same time as the transducer layers for TDTR samples. Using the Wiedemann-Franz Law, the thermal conductivity of the transducer layer is estimated to be ~94 W m$^{-1}$ K$^{-1}$. Since all the samples are prepared on the same wafer, we assume that they have the same Al transducer conductivity. The thermal interfacial resistance between the Al transducer layer and $SiO_2$ is measured using TDTR and is found to be ~4.4 m$^2$ K GW$^{-1}$ on a sample comprising Al [94 nm] / $SiO_2$ [157 nm] / Si. We assume that this thermal interfacial resistance also applies between the Al transducer and the nanostructured fin array layer in all of our samples. The thermal interfacial resistance between $SiO_2$ and Si is assumed to have a nominal value of 5 m$^2$ K GW$^{-1}$.[40] The volumetric specific heat of the fin layer is taken to be a 1:3 volume weighted average of Al and $SiO_2$. This leaves the cross-plane thermal conductivity of the fin layer, $\kappa_{NFA}$ as the only unknown parameter, which is extracted by fitting the data to the 3D heat diffusion model. In a later section, we calculate the performance of hypothetical, highly-scaled fin-arrays where quasi-ballistic effects within the Al might become more important. Error bars in $\kappa_{NFA}$ are predominantly based on uncertainties in the thickness of the nanostructured fin array layer, as measured by transmission electron microscopy (TEM). They also include the impact of aluminum oxidation on the interface between Al pillars and the top Al transducer and on heat capacity of Al transducer layer. (See Supplementary Information for uncertainty propagation and sensitivity analysis.)

The experimental results show that $\kappa_{NFA}$ increases with decreasing pitch of the microfabricated pillars, as shown in Fig. 2(b). Specifically, $\kappa_{NFA}$ for $L_P$ = 200 nm is ~1.7 times that for $L_P$ = 1.6 $\mu$m, more than two times the reference value with the same volume fraction of Al:$SiO_2$. This is mainly due to the increased contact area between Al and $SiO_2$ as the pitch $L_P$ decreases. To understand the physical mechanisms responsible for this enhancement in effective conductance of the nanostructured fin array layer, we analyze the various thermal conduction channels through the sample stack. The heat injected into the Al transducer predominantly flows into the pillars due to the much smaller thermal resistance of the Al pillars as compared to the surrounding $SiO_2$. The heat conducts three-dimensionally across the structured interface in the cross-plane (normal to the interface) and in-plane (parallel to the interface) directions. The in-plane heat flow contributes to the increase in the effective cross-plane thermal conductance of the nanostructured fin array layer as pitch $L_P$ decreases.

**SIMULATION OF NANOSTRUCTURED FIN ARRAY**

We use finite element numerical simulations to further investigate thermal conduction in the nanostructured fin array. The effective thermal conductivity of the fin layer is calculated by solving the heat diffusion equation within a geometry shown in Fig. 1 (see Supplementary Information for the details of the finite element simulations). To consider the size effects on the Al pillars, we treat Al pillars and the Al transducer layer to have separate thermal conductivity, and the value for the pillars is obtained using a model based on the kinetic theory as discussed below. We use the same thermophysical properties for the other materials and interfaces as used in the TDTR model shown in Fig. 2(a).

**Modeling Thermal Conductivity of Nanostructured Fin**

We use a model based on the kinetic theory to estimate the reduction in the thermal conductivity of Al pillars with decreasing pitch. The thermal conductivity of Al is the summation of the electron and phonon thermal conductivities, and the bulk thermal conductivity of each carrier can be written,

$$\kappa_{Al,bulk} = \sum_s \int_0^\infty \frac{1}{3} Cv \Lambda_{bulk} \, d\omega \tag{1}$$

where $C, v$, and $\Lambda_{bulk}$ are the volumetric heat capacity, the velocity, and the mean free path (MFP) of energy carrier in bulk Al, respectively. $s$ is band/mode index, and $\omega$ is angular frequency. As the pitch $L_P$ decreases, there is an increase in boundary scattering of the energy carriers within the Al pillars; this effect becomes significant when the feature size approaches the MFP of bulk Al $\Lambda_{Bulk}$. The MFPs of the electrons range from 1 nm to 20 nm at 300 K, and the MFPs of the phonons span from ~1 nm to ~8 nm at 300 K [41]. The Al pillar thermal conductivity of each carrier is written,

$$\kappa_{Pillar} = \sum_s \int_0^\infty \frac{1}{3} Cv \Lambda_{Pillar} \, d\omega \tag{2}$$

By changing variables from $\omega$ to $\Lambda_{bulk}$, Eq.(2) becomes [42]

$$\kappa_{Pillar} = \int_0^\infty \left[ -\sum_s \frac{1}{3} Cv \Lambda_{bulk} \left( \frac{d\Lambda_{bulk}}{d\omega} \right)^{-1} \right] \frac{\Lambda_{Pillar}}{\Lambda_{bulk}} d\Lambda_{bulk} \tag{3}$$

where the term in square brackets is the contribution of each MFP to thermal conductivity. The MFPs of Al pillar $\Lambda_{Pillar}$ is calculated using Matthiessen's rule, and the MFP is given by

$$\Lambda_{Pillar}^{-1} = d_{grain}^{-1} + \Lambda_{bulk}^{-1} + \Lambda_{boundary}^{-1} \tag{4}$$

where $d_{grain}$ is the averaged grain size, $\Lambda_{bulk}$ is the MFP of bulk Al, and $\Lambda_{boundary}$ is the MFP caused by boundary scattering. We obtain the MFP distribution for bulk Al and its contribution to the thermal conductivity from Jain *et al.* [41]. We use the measured Al transducer thermal conductivity to estimate the averaged grain size $d_{grain}$ to be ~12 nm, which is within a typical range of grain sizes in an evaporated Al film [43, 44]. We note that the averaged grain boundary can be larger than the estimated grain size since the value includes possible impact of impurity and imperfections. We assume that the grain size in the Al pillar is the same as in the transducer since the metallization conditions for the transducer and the pillars are identical. The MFPs due to boundary scattering, $\Lambda_{boundary}$, is calculated using Monte-Carlo approach. [45, 46] The thermal conductivity of Al pillars is estimated to be suppressed from ~92 Wm$^{-1}$K$^{-1}$ to ~80 Wm$^{-1}$K$^{-1}$ upon reducing the half pitch from 1 $\mu$m to 100 nm with 25% of Al volume fraction, when the thickness of pillars is fixed to 65 nm. We note that the ballistic effect on Al pillars becomes significant when $L_P$ is smaller than 200 nm. (See the Supporting Information for details of calculations.)

**THERMAL CONDUCTION MECHANISMS IN NANOSTRUCTURED INTERFACES**

To identify the various thermal conduction pathways and determine how the resistance of each pathway changes with $L_P$, we conduct simulations for four different cases: sidewall thermal boundary resistances of 0, 5, 20 m$^2$K GW$^{-1}$, and the limit of infinite thermal boundary resistance. Besides thermal conduction normal to the plane, three different planar conduction pathways are involved, namely, constriction in the top Al, spreading in the SiO$_2$ fin array below, and conduction through sidewall interfaces within the nanostructured fin array as shown in Fig. 3(a). The contribution of each pathway to the net thermal conduction is a function of $L_P$. For instance, with decreasing $L_P$, the reduced travel distance for heat flux in the lateral direction leads to a decrease in the constriction and spreading resistances. The effect is an increase in $\kappa_{NFA}$ with decreasing pitch for the case of infinite sidewall thermal resistance of sidewall interfaces as shown in Fig. 2 (b).

The measured effective thermal conductivity of the nanostructured fin array layer is still higher than the computed $\kappa_{NFA}$ value with infinite sidewall interfacial resistances. This difference is attributed to the impact of thermal conduction through sidewall interfaces. The effective conductivity $\kappa_{NFA}$ depends on the thermal resistance of sidewall interfaces between the Al pillar and the SiO$_2$ in the fin layer, and the sensitivity increases with decreasing pitch. We note that thermal conduction through sidewall interfaces becomes a major conduction path, which rarely plays an important role in interfacial thermal conduction.

By comparing our experimental data with finite element simulations, we estimate that the thermal resistance of sidewall interfaces lies in the range 5 to 20 m$^2$ K GW$^{-1}$. This resistance of the sidewall interface is higher than the experimentally measured value of the Al-SiO$_2$ planar interface (4.4 m$^2$ K GW$^{-1}$) possibly due to anisotropic etching and directional metallization; however, the value is still within the typical range for many metal-dielectric interfaces [9, 47-49]. We note that these experimentally measured values are often larger than the theoretical predictions based purely on acoustic mismatch and phonon-electron coupling, mainly due to extrinsic factors such as near interfacial defects, roughness and adsorbed impurities. As $L_P$ decreases, thermal conduction through sidewall interfaces dominates the temperature distribution in the SiO$_2$ region of the nanostructured fin array. Fig. 3(a) shows the temperature distribution in cross section of the unit cells for $L_P$ = 1 $\mu$m and $L_P$ = 200 nm. For $L_P$ = 200 nm, the lateral temperature gradient extends within the entire SiO$_2$ region between two adjacent fins, while for $L_P$ = 1 $\mu$m, it is limited to the vicinity of the fin. This indicates that the embedding material of the nanostructured fin array layer, which is SiO$_2$ in this work, increasingly participates in cross-plane heat transfer as pitch decreases.

**Thermal Conduction through Sidewall Interfaces**

Among the different thermal conduction pathways, transport across sidewall interfaces becomes a major conduction channel in the nanostructured interfaces as the pitch decreases. We estimate the thermal resistance offered to heat flow across the sidewalls of the Al pillars by analyzing the distribution of temperature and its gradient in their vicinity. This provides insights for relevant length scales of in-plane conduction, which can be useful in interface design. The mismatch in thermal conductivity between Al and SiO$_2$ works together with the local interface resistance and the geometry to induce lateral conduction. The contrasting thermal conductivities of the two materials, in this case Al and SiO$_2$ ($\kappa_{Al}$ = 94 W m$^{-1}$ K$^{-1}$ and $\kappa_{SiO_2}$ = 1.38 W m$^{-1}$ K$^{-1}$), sets up a temperature profile that exponentially decays in the in-plane direction with a characteristic length, $L_C$ [33]. We build a simplified thermal circuit model for the nanostructured fin array based on the layered stack configuration in Fig. 2(b), and the circuit is shown in Fig. 3(c). We note that constriction resistance in the Al can be neglected compared to spreading resistance in the SiO$_2$ since the thermal conductivity of Al is ~ 70 times higher than SiO$_2$.. Thermal resistance that is presented to heat flux going through sidewall interfaces is decoupled into the boundary resistance, $TBR"_{SW}$, and the resistance of SiO$_2$ in the pillar layer, denoted as $R"_{SW}$. The resistance $R"_{SW}$ is approximately,

$$R"_{SW} \cong \frac{2L_C}{\kappa_{SiO_2}} \tag{5}$$

The characteristic length $L_c$, known as the healing length along the layer of the nanostructured array, is $\sim\sqrt{t_{NFA}t_{SiO2}/2}$, where $t_{NFA}$ and $t_{SiO2}$ are the thicknesses of the fin-array layer and contiguous SiO$_2$ layers, respectively [29, 33]. (see Supplementary Information for details of thermal circuit derivation and analysis.) For the geometries used in this work, $L_C \approx$ 50 nm. We note that $L_C$ should be smaller than ~ $L_P$/12 to have most of the temperature rise decay within the unit cell. When $L_P$ is smaller than 600 nm, $L_C$ is geometrically limited by the pitch, and is therefore $\approx L_P$/12. In this regime, $R"_{SW}$ becomes proportional to the pitch $L_P$, and the resistance with $L_P$ = 200 nm is ~24 m² K GW⁻¹, which is comparable to $TBR''_{SW}$. With decreasing pitch, the boundary resistance $TBR''_{SW}$ plays an increasingly important role in the conduction pathway across sidewall interfaces.

**DESIGN GUIDELINES FOR NANOSTRUCTURED INTERFACES**

We further explore the impact of the aspect ratio and pitch from diffusive to quasi-ballistic transport regime in Al pillars, and this provides design guidelines for nanostructured interfaces. We calculate $\kappa_{NFA}$ varying the aspect ratio $t_{Pillar}/w_{Pillar}$ from 0.1 to 10 with $L_P$ = 40 nm, 200 nm, and 1.0 $\mu$m, where $t_{Pillar}$ and $w_{Pillar}$ are the thickness and the width of the Al pillars, respectively, as shown in the inset of Fig. 4. The reduction in thermal conductivity of the Al pillars is calculated using a model based on the kinetic theory (see inset of Fig. 4). The result of finite element calculations with pitch-dependent and constant Al thermal conductivity $\kappa_{Pillar}$, are shown with solid and dashed lines, respectively. The impact of the aspect ratio can be grouped into two regimes: $t_{Pillar}/w_{Pillar} < 1$ and $t_{Pillar}/w_{Pillar} > 1$. When $t_{Pillar}/w_{Pillar} < 1$, $\kappa_{NFA}$ increases very slowly with aspect ratio as shown in Fig. 4, since the interfacial resistance of sidewall interfaces $TBR''_{SW}$ is larger than the resistance of other conduction paths. On the other hand, when $t_{Pillar}/w_{Pillar} > 1$, $\kappa_{NFA}$ increases rapidly with increasing aspect ratio, mainly due to the increased thermal conduction through sidewall interfaces. In this regime, the impact of the reduction in $\kappa_{Pillar}$ becomes pronounced, as shown by the difference between solid and dashed lines in Fig. 4. Even though the reduced thermal conductivity may negatively impact thermal conduction over the fin array, the enhancement due to increased contact area is beneficial for any aspect ratio. We note that $\kappa_{NFA}$ collapses onto a single curve for the case of constant $\kappa_{Pillar}$ when $t_{Pillar}/w_{Pillar} > 1$ while such convergence is not observed when $t_{Pillar}/w_{Pillar} < 1$ due to the non-negligible spreading and constriction resistances. This analysis suggests that the concept of nanostructured interfaces can be extended to a combination of a diffusive medium and a quasi-ballistic medium despite a reduction in the thermal conductivity of the quasi-ballistic medium.

We note that if this idea must be extended to interfaces between two materials which both have long intrinsic mean free paths compared to the feature size, additional factors will need to be considered in the design process. While nanostructuring the interface through fin-like projections will increase the thermal conductance, decreasing feature sizes could also lead to lower near-interfacial thermal conductivities in the adjacent materials. This suggests an optimum pitch where the total thermal conductance is maximized, when the competing effects of increased contact area and phonon size-effects balance each other out. This merits further study and is the focus of ongoing work.

**CONCLUDING REMARKS**

In the present work, we show that the solid-state nanostructured fin array is a new and promising approach to lower the thermal boundary resistance. We demonstrate that the nanostructured fin array with $L_P = 200$ nm enhances interfacial thermal conductance by more than a factor of two compared to an equivalent volume of stacked materials with a planar interface. The enhanced thermal conduction of the nanostructured fin array is immediately applicable to the design of near-junction layers for thermally limited electronics such as integrated nanoelectronics and silicon-on-insulator (SOI) devices. Numerical simulations and theoretical calculations identify the contribution of the various conduction mechanisms in the nanostructured fin array, namely constriction and spreading resistances and heat conduction across sidewalls. This highlights the role of sidewall interfaces in thermal transport for highly integrated nano-systems. We use a reduced-order analysis, which provides design guidelines for solid-solid interfaces. This simplified reduced-order model can also be applied to the complex simulation of many practical nanostructures, such as interconnects in microelectronics. We further discuss the impact of quasi-ballistic effects within the Al pillars in hypothetical, highly-scaled fin-arrays, and show that the nanostructured fin array remains an effective solution in this regime.

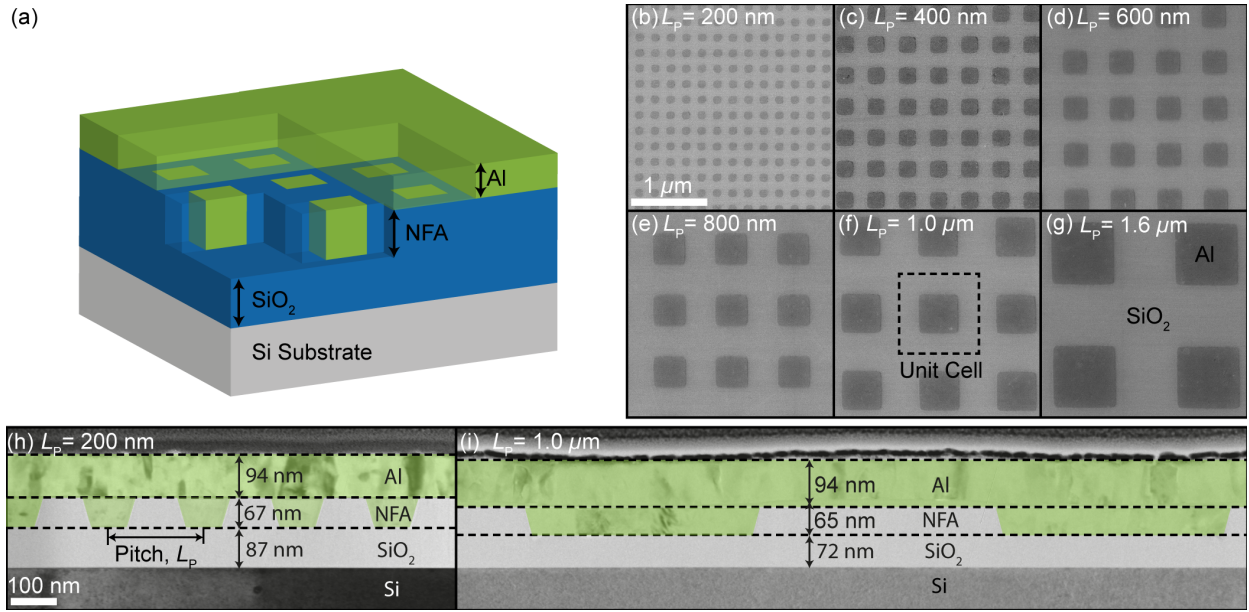

**Figure 1.** (a) Schematic of nanostructured fin array (NFA). (b-g) Plan-view scanning electron microscopy (SEM) images for samples with different pitch $L_P$, taken before deposition of Al transducer layer. (h, i) Cross-sectional transmission electron microscopy (TEM) images for samples with $L_P$ = 200 nm and $L_P$ = 1.0 $\mu$m. The Al regions in the nanostructured fin array and the top transducer layers are false-colored in green. (b-g) and (h-i) share scale bars shown on (b) and (h), respectively.

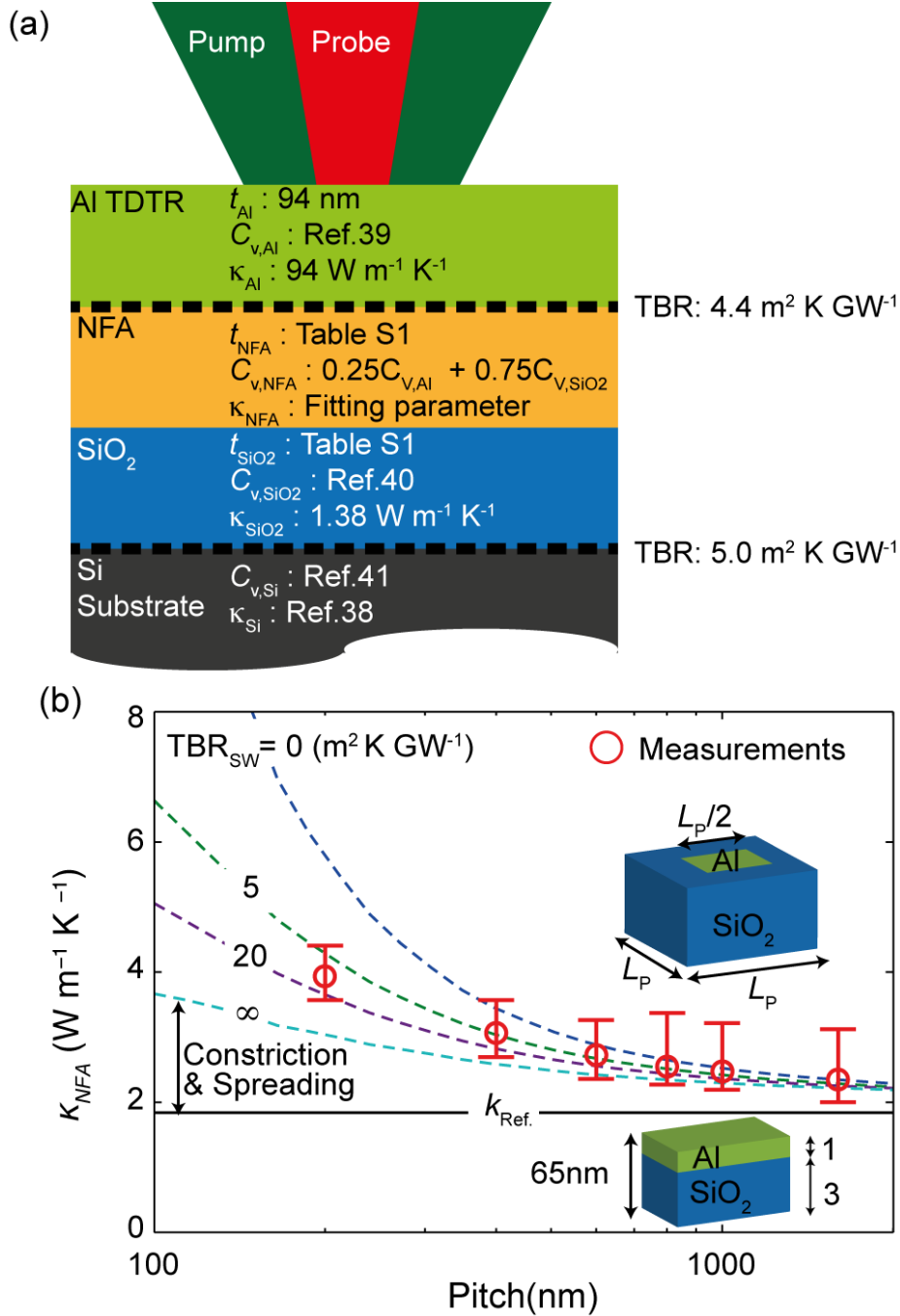

**Figure 2**. (a) Schematic of sample stack used to analyze time-domain thermoreflectance data. The nanostructured fin array (NFA) is assumed to be a homogeneous layer. $t$, $C_V$, and $\kappa$ denote thickness, volumetric heat capacity, and thermal conductivity of a layer, respectively. Dashed lines indicate interfaces with finite resistance (b) Effective cross-plane thermal conductivity of the nanostructured fin array layer, $\kappa_{NFA}$, as a function of the pitch, $L_\text{p}$. The measurement results are shown as red circles with error bars, and the dashed lines are obtained from the finite-element simulations with different values of the sidewall thermal boundary resistance. The black solid line indicates the reference case when the samples have a layered structure as shown in the bottom-right of the plot.

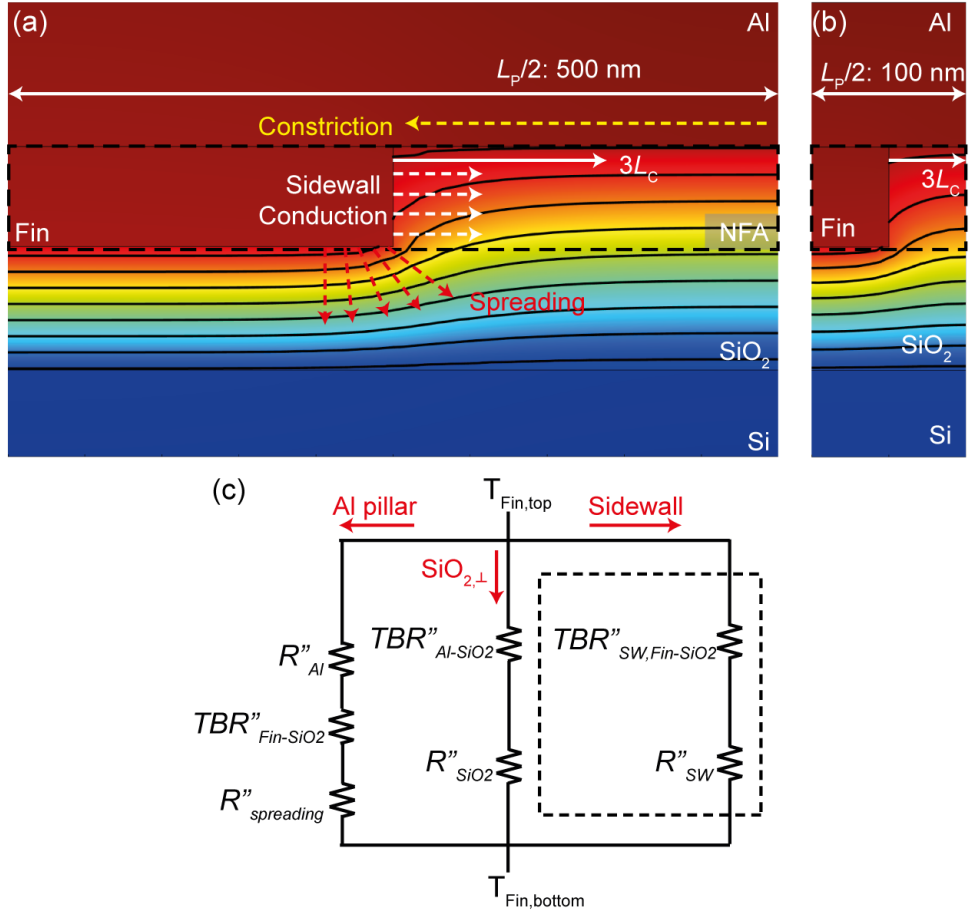

**Figure 3.** (a,b) Temperature profiles corresponding to $L_P = 1$ μm and $L_P = 200$ nm, calculated using the finite-element method. The black solid and dashed lines are isotherms and the boundary of the nanostructured fin array (NFA), respectively. (a) The dashed arrows show the thermal conduction mechanisms within the stack. (c) Simplified thermal circuit model for the system showing the various thermal pathways including spreading resistance, across $SiO_2$ in the layer of nanostructured fin array, and sidewall interface. The subscripts Al and Fin indicate the transducer and metal embedded in $SiO_2$, respectively, and SW indicates sidewall. $T_{Fin,top}$ and $T_{Fin,bottom}$ are the averaged temperatures on the top and bottom boundaries of the fin layer.

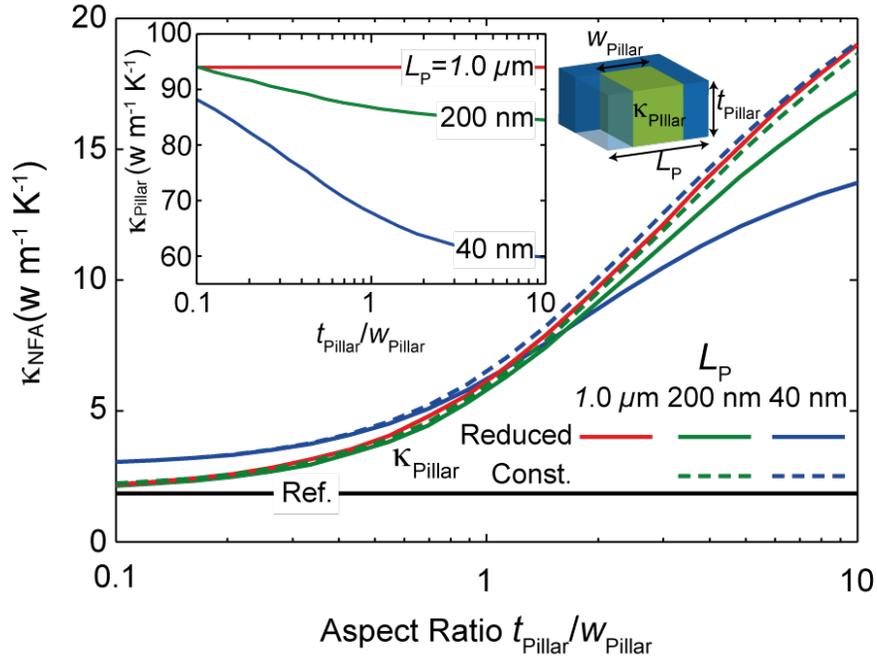

**Figure 4**. The effective thermal conductivity of the nanostructured fin array (NFA) $\kappa_{NFA}$ with varying aspect ratio (defined as the ratio of the pillar thickness, $t_{Pillar}$ to the pillar width, $w_{Pillar}$). The Al:SiO$_2$ fill fraction in all cases is 25%, and $TBR''_{SW}$ is set to 4.4 m$^2$ K GW$^{-1}$ for all interfaces between Al and SiO$_2$. Solid and dashed lines denote $\kappa_{NFA}$ calculated with pitch-dependent and constant Al thermal conductivity ($\kappa_{Pillar}$), respectively. The inset shows $\kappa_{Pillar}$ as a function of aspect ratio ranging from 0.1 to 10. The red, green, and blue colors correspond to pitch $L_P$ = 1000 nm, 200 nm, and 40 nm, respectively. The black solid line indicates the reference layered structure as shown in Fig. 2(b).

## NOMENCLATURE

| | |
|---|---|
| $C$ | heat capacity |
| $d_{grain}$ | grain size |
| $L_C$ | healing length along the layer of the nanostructured array |
| $L_P$ | pillar-to-pillar pitch |
| $R"$ | volumetric thermal resistance |
| $TBR"$ | thermal boundary resistance |
| $s$ | band/mode index |
| $t_{Pillar}$ | thickness of pillar |
| $v$ | velocity of energy carriers |
| $V_{in}$ | in-phase voltage signal on lock-in amplifier |
| $V_{out}$ | out-of-phase voltage signal on lock-in amplifier |
| $w_{Pillar}$ | width of Al pillar |

**Greek Symbols**

| | |
|---|---|
| $\kappa$ | thermal conductivity |
| $\kappa_{Al}$ | thermal conductivity of Al transducer, ~94 W m$^{-1}$ K$^{-1}$ |
| $\kappa_{NFA}$ | thermal conductivity of a nanostructured fin array in cross-plane direction |
| $\kappa_{Pillar}$ | thermal conductivity of Al pillar |
| $\kappa_{SiO_2}$ | thermal conductivity of SiO$_2$, 1.38 W m$^{-1}$ K$^{-1}$ |
| $\Lambda$ | mean free path of energy carriers |
| $\omega$ | angular frequency of phonon |

**Subscripts**

| | |
|---|---|
| $boundary$ | due to boundary scattering |
| $bulk$ | due to scattering mechanisms in bulk Al |
| $SW$ | sidewall interfaces |

**Supplementary Information**

Additional information on sample fabrication, sample dimensions, experimental details of time domain thermoreflectance (TDTR), TDTR sensitivity and uncertainty analysis, details of finite element simulations, estimation of grain size, mean free paths due to boundary scattering in Al pillar, and mathematical details of reduced order analysis.

## AUTHOR INFORMATION

**Corresponding Author**

*E-mail: goodson@stanford.edu.

## ACKNOWLEDGMENT

This work is supported by Semiconductor Research Corporates (Agreement No. 2012-OJ-2308). The authors acknowledge the use of the Stanford Nano Shared Facilities (SNSF) of Stanford University for sample preparation and characterization.

1. Sample Fabrication

We pattern the square holes for Al pillars using electron-beam lithography on top of ~157 nm thick thermally grown $SiO_2$ layer on Si substrate. Reactive ion etching (RIE) is used to etch holes with square cross-section in the $SiO_2$ up to a depth of ~92 nm. The holes are filled with Al using electron-beam evaporation, following which the surface is planarized back down to the top of the $SiO_2$ layer using chemical mechanical polishing (CMP). $SiO_2$ surrounding the Al pillars acts as a CMP stop since it is chemically resistive to the components in the CMP slurry. Finally, a ~94 nm thick Al layer is deposited on top of the planarized surface to serve as an opto-thermal transducer layer for thermoreflectance measurements. On the same wafer, non-etched $SiO_2$ samples are prepared for the thermal characterization of the Al transducer layer and its interface with $SiO_2$. The pillar-to-pillar pitch is confirmed using plan-view scanning electron microscopy (SEM) before the Al transducer is deposited. Thicknesses of the various layers, namely the Al transducer, nanostructured fin array and buried $SiO_2$ are measured using cross-sectional transmission electron microscopy (TEM) (see the section 2 for details), and the thicknesses of $SiO_2$ both before and after etching are confirmed using optical ellipsometry.

2. Sample dimensions

We use cross-sectional transmission electron microscopy (TEM) to determine the thicknesses of individual layers for all the samples (see Fig. 1 (g, h) in the main manuscript and Fig. S1 below). The thickness of the $SiO_2$ layer located below the nanostructured fin array (NFA) is defined as the distance from Si-$SiO_2$ interface to the flat region at the bottom of the Al fin. The dimensions for the $SiO_2$ layers increases with decreasing pitch $L_P$ since the etch rate for $SiO_2$ slows as area of a hole decreases. The thickness of the fin array layer is defined as the vertical distance from the flat region at the bottom of the Al fin to the edge of the trench, which corresponds to the height of sidewall cross-section. The upper and lower bounds are measured from the base of the fin layer to the highest point of the $SiO_2$ in the layer of the fin array, and the lowest point of the interface between the Al pillar and the transducer layer, respectively. This variation in film thicknesses is mainly attributed to the planarization process in fabrication. The fluctuation reduces with decreasing pitch $L_P$ since the $SiO_2$ in the fin layer serves as a more effective CMP stop for material in its proximity. The measured thicknesses are summarized in Table S1.

**Table S1.** Layer thicknesses determined by cross-sectional TEM

| Layer Thickness | | Pitch $L_P$ | | | | | |
|---|---|---|---|---|---|---|---|
| | | 200 nm | 400 nm | 600 nm | 800 nm | 1.0 µm | 1.6 µm |
| NFA Layer (nm) | Upper Bound | 69 | 72 | 71 | 68 | 73 | 75 |
| | Nominal | 67 | 63 | 67 | 61 | 65 | 66 |
| | Lower Bound | 65 | 61 | 62 | 58 | 62 | 60 |
| $SiO_2$ Layer (nm) | | 87 | 78 | 75 | 73 | 72 | 68 |

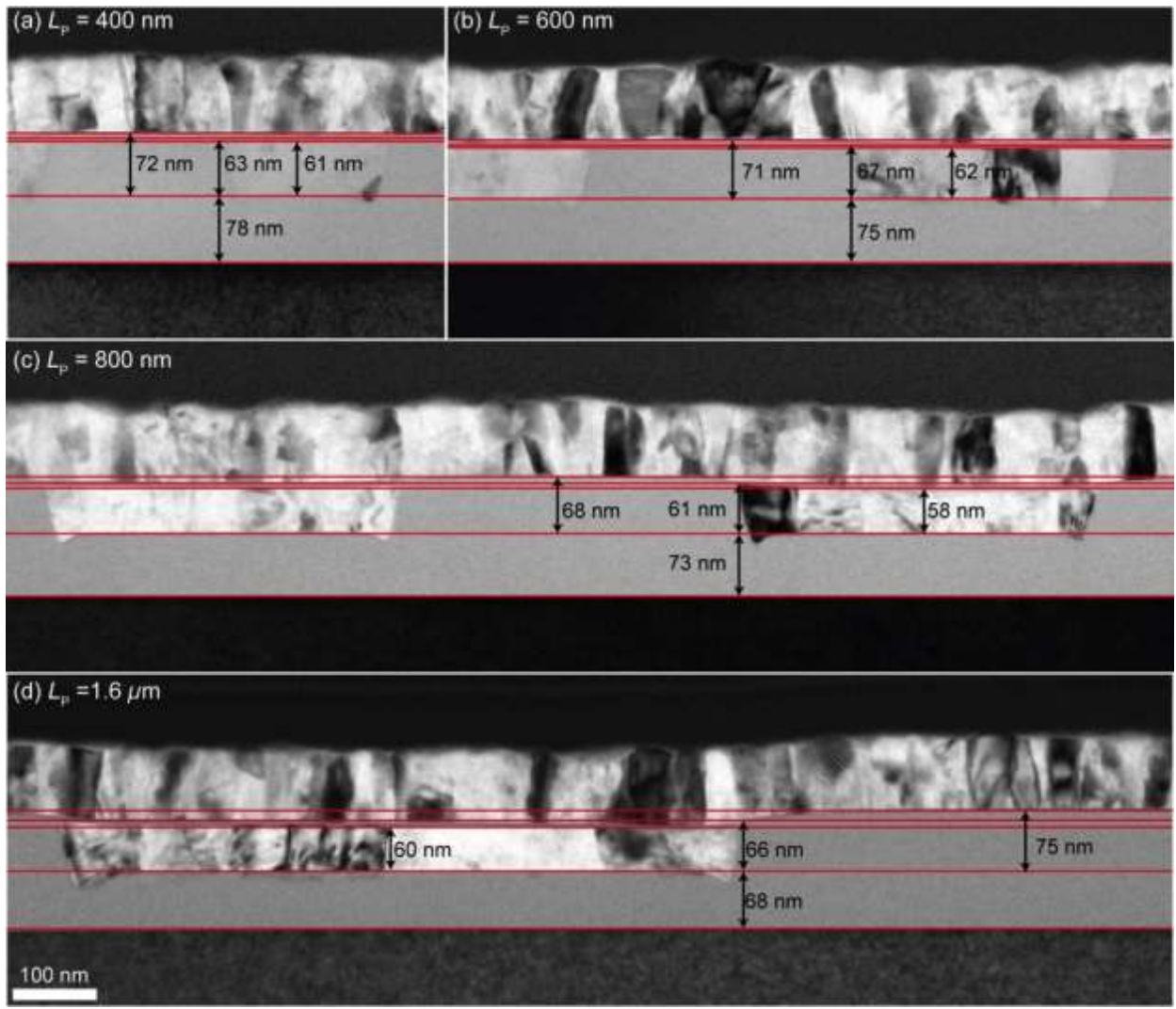

**Figure S1.** Cross-section TEMs for the samples not shown in the main text. All of the images share the scale bar shown in (d). The thickness of the fin array (NFA) is shown with both the upper and lower bounds.

### 3. Time Domain Thermoreflectance (TDTR)

We measure the effective thermal conductivity of the nanostructured fin array layer in the film-normal direction, $\kappa_{NFA}$, using time-domain thermoreflectance (TDTR), an optical pump-probe technique [1-3]. TDTR technique utilizes ~ 9 ps optical pulses from a pump source to heat up the surface of a sample coated with a thin Al transducer layer, and time-delayed probe pulses to interrogate the temporal changes in transducer reflectivity as heat diffuses into the sample of interest. In these experiments, pump pulses are amplitude modulated at a frequency of 4 MHz to enable lock-in detection. Laser spot sizes (1/e² diameters) of 10.2 ± 0.1 $\mu$m and 6.2 ± 0.2 $\mu$m are used for pump and probe beams, respectively. The total input optical power is ~10 mW, resulting in a steady-state temperature rise of ~2 K. The experimental data consist of the in-phase ($V_{in}$) and out-of-phase ($V_{out}$) voltage signals generated by the reflected probe intensity at the lock-in amplifier, as a function of time-delay from 0 to 3.6 ns. Unknown thermal properties in the sample stack are extracted by fitting the time-series of the ratio (-$V_{in}$/$V_{out}$) signal to a three-dimensional solution of the heat equation, accounting for finite conductance of interfaces, Gaussian distribution of optical intensity, and in-plane thermal spreading effects [4].

### 4. Sensitivity Analysis

To quantify the sensitivity of the TDTR measurements to the various thermophysical parameters in the sample stack, we define and calculate the sensitivity coefficients using the equation below:

$$S_\alpha = \frac{\partial \log\left(-\frac{V_{in}}{V_{out}}\right)}{\partial \log(\alpha)} \tag{S12}$$

where $\alpha$ is the parameter of interest, $V_{in}$ and $V_{out}$ are respectively the in-phase and out-of-phase voltage signals at a given time delay. The sensitivity $S_\alpha$ variation with time delay for some important parameters is shown in Fig. S2. A value close to zero indicates negligible sensitivity, while a large absolute value implies good sensitivity. The measurements are insensitive to the in-plane conductivity of the fin-array layer $\kappa_{Fin,\parallel}$ as shown in Fig. S2, because the thermal penetration depth at the modulation frequency of 4 MHz (~250 nm) is much smaller than the spot size (~10 µm). Further, we have minimal sensitivity to the thermal boundary resistances (TBRs) at the Al/Fin, Fin/SiO$_2$ and SiO$_2$/Si interfaces. We use cross-sectional transmission electron microscopy to accurately determine the thicknesses of various layers, namely the top Al transducer, the fin-array and the SiO$_2$, since the measurements are sensitive to these dimensions. Oxidation of aluminum poses additional sources of uncertainty: 1) additional interfacial resistance between the pillars and the transducer and 2) variation in heat capacity of the transducer layer. The impact of aluminum oxidation on the measurements is summarized in Measurement Uncertainty section.

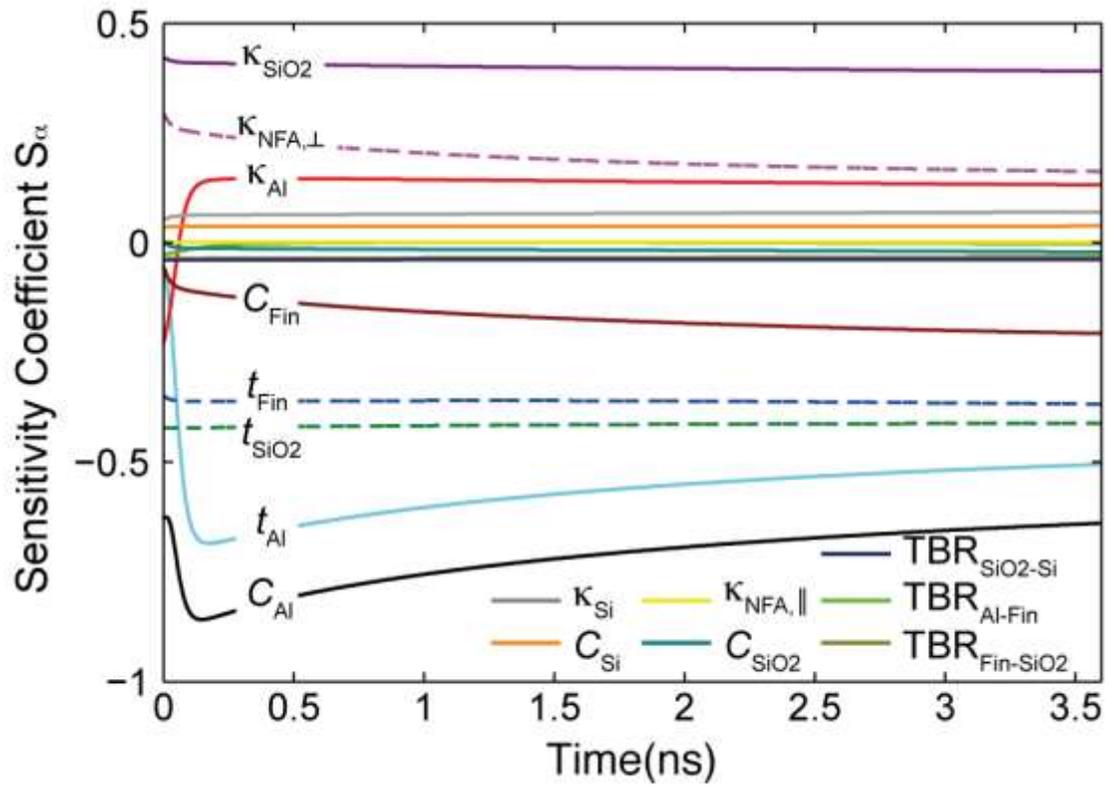

**Figure S2.** Sensitivity analysis for all parameters of the nanostructured fin array (NFA).

## 5. Measurement Uncertainty

We quantify uncertainties in the extracted conductivities by perturbing the different input fitting parameters based on their individual uncertainties. The predominant source of error is the variation in the thickness of the layer of the nanostructured fin array (NFA), $t_{NFA}$ due to inherent unevenness introduced because of the fabrication process. We use cross-sectional TEM images to determine the upper and lower bounds of $t_{NFA}$ as shown in Table S1, and the propagated uncertainties for all samples are summarized in Table S2.

Besides the predominant uncertainty due to variation in dimensions, Al can be oxidized at the interface between Al pillars and the Al transducer layer and on top of the transducer layer, creating additional sources of measurement uncertainty. To quantify the impact of the native oxide between the pillars and the transducer, we assume a ~3 nm thick native oxide on top of Al pillars[5] which can be treated as an interfacial resistance. The native oxide between an Al pillar and the top transducer layer occupies a fourth of the unit cell area, and its thermal conductivity is ~1.6 Wm$^{-1}$K$^{-1}$.[6] The thermal boundary resistance due to the oxide is estimated to be ~0.4 m$^2$KGW$^{-1}$, which causes a variation of < 1 % in the NFA thermal conductivity. We further investigate the impact of the oxidation of Al transducer on its volumetric specific heat. Assuming a ~3 nm thick layer of Al$_2$O$_3$, the volumetric specific heat of the transducer layer increases by ~0.7 %[7, 8]. This results in up to ~8% variation in thermal conductivity of NFA. We note that presence of a native oxide causes marginal change in experimental results, and the propagated uncertainties due to Al oxidation are summarized in Table S2.

**Table S2.** Error propagation analysis for TDTR measurements

| Pitch $L_P$ (nm) | Thickness of NFA | | | | TBR Al pillar - Al transducer | $C_{Al}$ (~0.7%) | Total | |
|---|---|---|---|---|---|---|---|---|
| | $t_{NFA}$ (nm) | | $\kappa_{NFA}$ (Wm$^{-1}$K$^{-1}$) | | $\kappa_{NFA}$ (Wm$^{-1}$K$^{-1}$) | $\kappa_{NFA}$ (Wm$^{-1}$K$^{-1}$) | $\kappa_{NFA}$ (%) | |
| | Error (-) | Error (+) | Error (-) | Error (+) | Error (-/+) | Error (+) | Error (-) | Error (+) |
| 200 | 2 | 2 | 0.37 | 0.41 | -0.033 | 0.33 | 9.4 | 13.5 |
| 400 | 2 | 9 | 0.37 | 0.49 | -0.007 | 0.18 | 12.2 | 16.9 |
| 600 | 5 | 4 | 0.37 | 0.53 | -0.001 | 0.14 | 13.6 | 20.0 |
| 800 | 3 | 7 | 0.27 | 0.82 | 0.004 | 0.13 | 10.7 | 32.7 |
| 1000 | 3 | 8 | 0.28 | 0.74 | 0.003 | 0.11 | 11.4 | 30.2 |
| 1600 | 6 | 9 | 0.35 | 0.77 | 0.005 | 0.09 | 15.0 | 33.2 |

## 6. Finite Element Simulations

We use finite element numerical simulations to further investigate thermal conduction in the nanostructured fin array. The computational domain comprises a single unit cell of the two-dimensionally periodic nanostructured fin array structure. Symmetry boundary conditions are applied on the sides, which are justified given the large heating spot size of the laser (~10 $\mu$m) used for TDTR, measurements relative to the pitch of our structures. A uniform heat flux is applied over the top Al surface, while the temperature at the bottom of the 10 $\mu$m thick Si substrate is fixed at 290 K, which is thick enough to ensure a uniform temperature distribution on the bottom plane. To consider size effects on the Al pillars, we treat Al pillars and the Al transducer layer to have separate thermal conductivity, and the value for the pillars is obtained using a model based on the Boltzmann transport equation as discussed in the main text. We use the same thermophysical properties for the other materials and interfaces as used in the TDTR model shown in Fig. 2(a). The average temperature on the top Al surface is computed by solving the Fourier heat diffusion equations using the COMSOL Multiphysics package. The effective thermal resistance of the system is computed using the calculated temperature difference between the top and bottom surfaces of the computational domain under the applied heat flux. The thermal resistance of the nanostructured fin array layer is extracted by subtracting the thermal resistances of the Al transducer, the $SiO_2$ layer below the fin structure, and the Si substrate from the total resistance.

## 7. Estimation of Grain Size

The average grain size of the Al transducer is estimated using its experimentally measured thermal conductivity. We use Eqs. (1)-(4), applying the same strategy for Al pillars to the reduction in thermal conductivity of Al transducer layer, and $\Lambda_{boundary}$ is approximated to be the thickness of the transducer layer. By fitting the thermal conductivity of the Al transducer, the averaged grain size $d_{grain}$ is estimated to be ~12 nm as shown in Fig. S3. This is within a typical range of grain sizes in an evaporated Al film [9, 10]. We note that the averaged grain boundary can be larger than the estimated grain size since the value includes possible impact of impurity and imperfections.

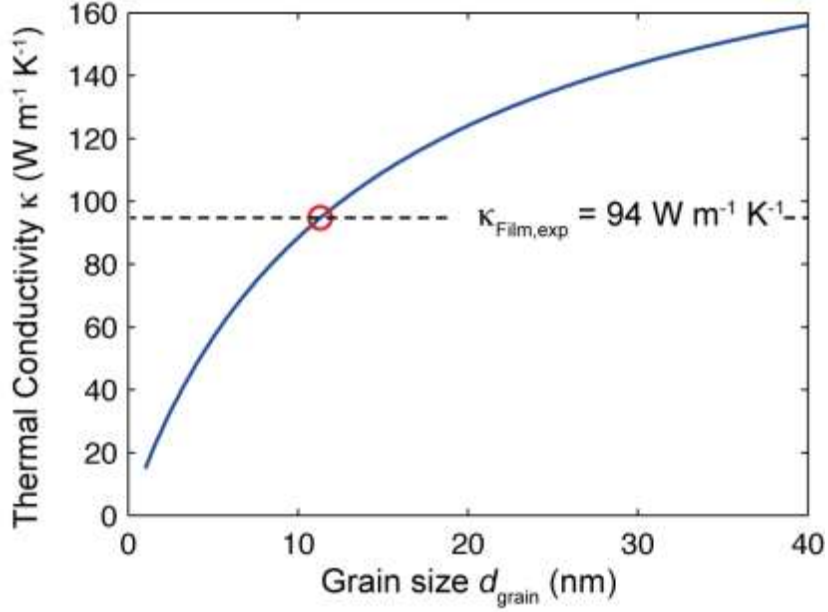

**Figure S3.** Thermal conductivity of Al depending on grain size. The dashed line is the experimentally measured thermal conductivity for the Al transducer, and the red marker indicates the grain size that matches the experimental thermal conductivity, ~ 12 nm.

8. **Mean Free Paths due to Boundary Scattering in Al Pillar**

We use a Monte-Carlo type approach to estimate $\Lambda_{boundary}$ in the pillars. To determine a free path due to boundary scattering, we simulate random particles with randomly selected locations and directions in a pillar with a boundary that is connected to the top Al transducer layer[11, 12]. A free path is calculated as the distance between a randomly selected origin and surrounding boundaries along a given direction. This calculation is repeated for 1000 times until average of the free paths converges, and the averaged value is $\Lambda_{boundary}$. The contribution of each MFP to thermal conductivity is suppressed by multiplying it with $(\Lambda_{Pillar}/\Lambda_{bulk})$ as seen in Eq. (3). The contribution of the MFP is summed to get the spatially averaged thermal conductivity value of the Al pillar. The thermal conductivity is estimated to be suppressed from

~92 Wm$^{-1}$K$^{-1}$ to ~44 Wm$^{-1}$K$^{-1}$ upon reducing the half pitch from 1 μm to 10 nm with 25% of Al volume fraction, when the thickness of pillars is fixed to 65 nm, as shown in Fig. S4. We note that the ballistic effect on Al pillars becomes significant when $L_P$ is smaller than 200 nm.

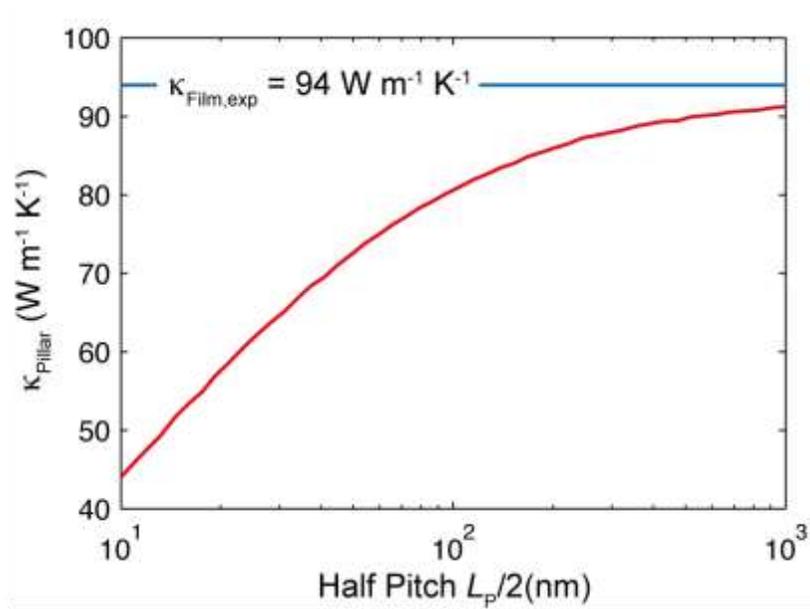

**Figure S4.** The effective thermal conductivity $\kappa_{Pillar}$ with varying half pitch $L_P/2$ from 10 nm to 1 μm, where the thickness of the Al pillar is fixed to be 65 nm.

## 9. Reduced order thermal model for nanostructured interfaces

We first derive a temperature distribution in the SiO$_2$ of a nanostructured fin array. Heat applied uniformly to the top of the transducer layer conducts two-dimensionally across the structured interface in the cross-plane and in-plane directions. Incoming heat flux to the SiO$_2$ region of the layer of nanostructured fin array is a combination of these heat flux, *i.e.*, transport through sidewall interfaces with Al pillars and planar interfaces with a Al transducer layer. This incoming heat flux dominates the temperature distribution in the SiO$_2$ region. The temperature profile in the SiO₂ is the combination of these impacts of heat flux, which can be approximated as,

$$T(r,z) \approx A(z)\exp\left(\frac{-(r-r_0)}{L_C}\right) + B(z) \tag{S1}$$

where $L_C$ is the characteristic thermal length-scale, often referred to as thermal heating length, within the SiO₂ region of the layer of the fin array. $r_0$ is the distance from the center of a pillar to its outer boundary, *i.e.*, the half of the pillar width, as shown in Fig. S5(a). The exponential term is due to heat conduction in the in-plane direction, while the second term describes conduction in the cross-plane direction in the SiO₂ of the fin array. To simplify the analysis, we assume the fin structure and its unit-cell have cylindrical symmetry. Fig. S5(a) shows a cross-sectional view of half of the unit cell. The characteristic thermal length-scale is estimated to be $L_c = \sqrt{t_1 t_2/2}$, which is ~ 50 nm in our case.[13, 14]

To satisfy periodic boundary conditions, the temperature gradient in the radial direction must be zero at the lateral boundaries of the unit cell. The temperature rise in radial direction has insufficient length to decay out within SiO₂ region of the fin array as pitch decreases. For $L_P <$ 600 nm, the characteristic thermal length-scale within SiO₂ is limited by the pitch. In this regime, we approximate the characteristic length to be $L_C = r_0/3$, in order to ensure ~ 95% of the temperature decays within the unit cell. Fig. S5(b) shows the normalized radial temperature profile within the SiO₂ region of the fin array at different planes across its thickness. The normalized temperature is defined as below:

$$T_{nor} = \frac{T(r,z) - B(z)}{A(z)} \tag{S2}$$

The normalized temperature curves fit well to exponential decay profiles, with the corresponding characteristic length-scales given by $L_c$.

We determine the closed form of the thermal resistance that is presented to heat flux going through sidewall interfaces using the approximate temperature profile derived above. The term *B(z)* in Eq. S1 is first written down under the assumption that it describes purely 1D conduction in the film-normal direction, between the Al transducer and the Si substrate, through the SiO$_2$ layer:

$$B(z) = \left(\frac{T_{fin,h} - T_{fin,c}}{t_2}\right) z + \left(-\frac{t_1}{t_2} T_{fin,h} + \left(1 + \frac{t_1}{t_2}\right) T_{fin,c}\right) \tag{S3}$$

where $T_{fin,h}$ and $T_{fin,c}$ denote $T(2r_0, t_1+t_2)$, and $T(2r_0, t_1)$, respectively as shown in Fig. S5. Using the temperature profile, the heat flux through sidewall interfaces calculated to be

$$q_{SW}(z) = -A_{SW} \kappa_{SiO_2} \left.\frac{\partial T}{\partial r}\right|_{r=r_0} = \frac{\kappa_{SiO_2} A_{SW}}{L_C} \frac{\Delta T(z)}{1 - \exp\left(-\frac{r_0}{L_C}\right)} \tag{S4}$$

where, $\Delta T(z) = T(r_0, z) - T(2r_0, z)$ and $A_{SW}$ is the sidewall area of the Al pillar. To simplify the analysis, $q_{SW}$ is averaged over the entire thickness of NFA, and $\Delta T(z)$ is replaced with $\Delta T_{avg}$,

$$\Delta T_{avg} = \frac{1}{t_2} \int_{t_1}^{t_1+t_2} T(r_0, z) - T(2r_0, z) dz \tag{S5}$$

We note that a good approximation is that $T(r_0,z) = T(r_0, t_1+t_2) = T_{fin,h}$, since Al is significantly more thermally conducting compared to SiO$_2$, and is therefore isothermal Eq. (S6-8) show how we simplify $\Delta T_{avg}$.

$$\Delta T_{avg} = \frac{1}{2}(T_{fin,h} - T_{fin,c}) = \frac{1}{2} \Delta T_{fin} \tag{S6}$$

$$\frac{1}{t_2} \int_{t_1}^{t_1+t_2} T(r_0, z) dz \cong T_{fin,h} \tag{S7}$$

$$\frac{1}{t_2}\int_{t_1}^{t_1+t_2} T(2r_0, z)dz \cong \frac{1}{2}\left(T_{fin,h} + T_{fin,c}\right) \tag{S8}$$

The heat flux through the sidewall interface, and the corresponding thermal resistance $R_{SW}$ can be written as follows,

$$q_{SW} = \kappa_{SiO_2} \frac{A_{SW}}{L_C} \frac{\frac{1}{2}\Delta T_{fin}}{1 - \exp\left(-\frac{r_0}{L_C}\right)} \tag{S9}$$

$$R_{SW} = \frac{2L_C}{\kappa_{SiO2} A_{SW}} \left(1 - \exp\left(-\frac{r_0}{L_C}\right)\right) \tag{S10}$$

We can further approximate $R_{SW}$ with 5% error as,

$$R_{SW} \cong \frac{2L_C}{\kappa_{SiO2} A_{SW}} \tag{S11}$$

Knowing the thermal resistance due to sidewall thermal conduction, the equivalent thermal circuit for the nanostructured fin array is simplified to that shown in Fig. 3 (b). We predict the effective thermal conductivity of the fin array using this thermal circuit model with thermal resistance of sidewall interface of 5 and 20 (m² K GW⁻¹). The thermal resistance for the other pathways is extracted from numerical simulations, under the assumption that there is no heat flux across the sidewall interfaces. The reduced order model prediction is shown in Fig. S6. Although this reduced order model deviates from the finite-element numerical calculations up to ~24%, the reduced order model provides insights on thermal conduction through sidewall interfaces. The discrepancy between the thermal circuit model and the numerical calculation is primarily attributed to the assumption that the entire volume of Al is isothermal.

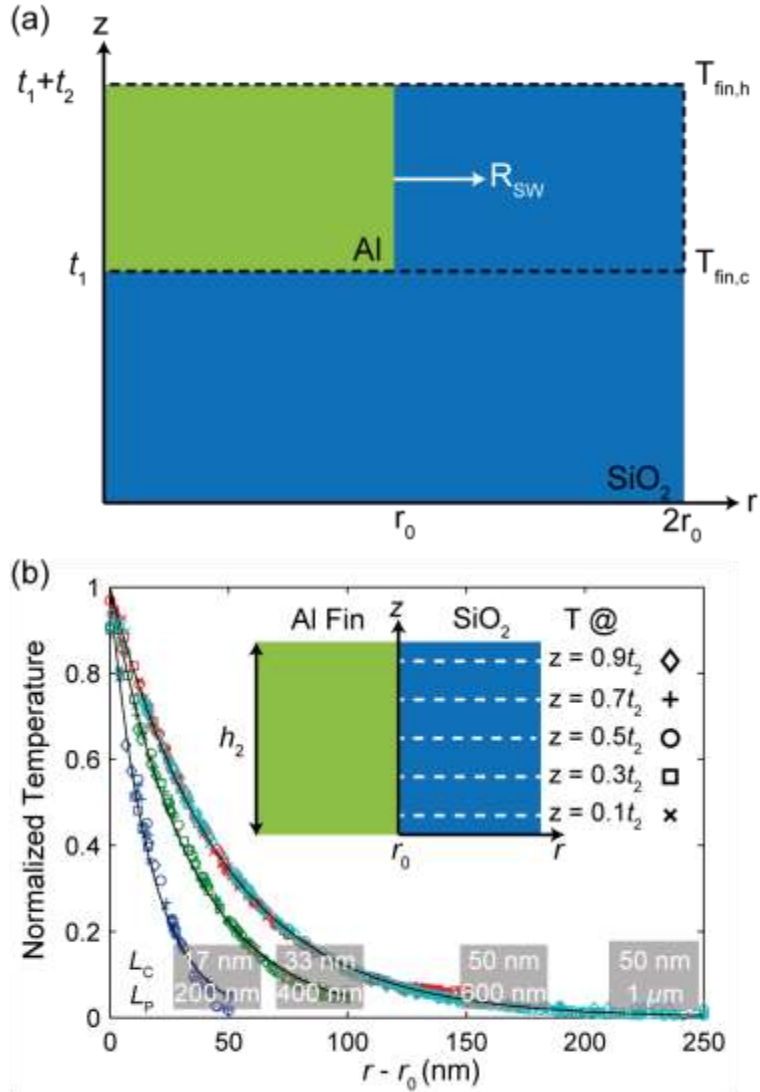

**Figure S5.** (a) Cross-sectional schematic of half of a fin and the contiguous $SiO_2$ layers drawn using cylindrical coordinates. The fin layer is outlined within the dashed lines. Fin radius $r_0$ is half of the radial distance from the fin-center to the adiabatic lateral surface. (b) Normalized radial temperature profile for NFA with various pitches with Eq. S2. The normalized profile is shown along planes that are at different heights within the thickness of the NFA layer as shown in the inset.

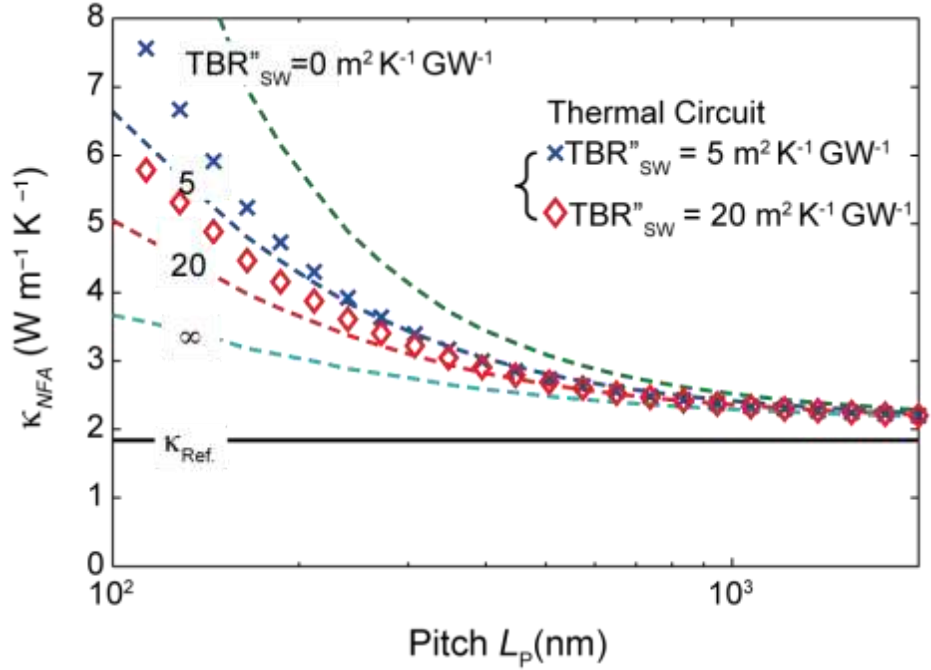

**Figure S6.** Thermal conductivity predictions for nanostructured fin array. The dashed lines are obtained from finite-element numerical simulations, while markers are from the reduced order model. The values for the reduced model are calculated using sidewall thermal boundary resistances ($TBR''_{SW}$) of 5 and 20 (m² K⁻¹ GW⁻¹), shown by the blue crosses and red diamonds, respectively.

## 10. Experimental Results in Conductance

We display the experimental data of nanostructured fin array (NFA) in terms of the thermal conductance, and the values are calculated for a 65 nm thick NFA.

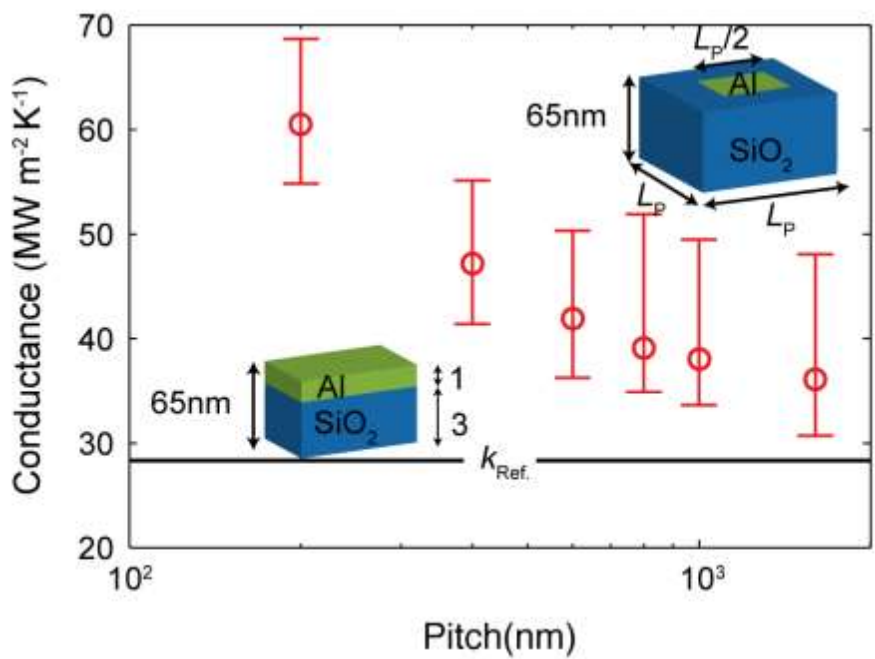

**Figure S7.** Effective cross-plane thermal conductance of the nanostructured fin array as a function of the pitch $L_P$. The black solid line indicates the reference case when the samples have a layered structure as shown in the bottom-left of the plot.